\documentclass[11pt]{article}
\usepackage[french,english]{babel}
\usepackage{amsmath}
\usepackage{amsfonts}
\usepackage{amssymb}
\usepackage{graphicx}   
\usepackage{times}
\def\hbar{/\ \hskip -9pt $h$}     
\def\psl {/\ \hskip -9pt p}       

\begin{document}
%
May 11\,th, 2009
\null\vskip 2.5cm
{\baselineskip 20pt
\begin{center}
{\bf LARGE MIXING ANGLES IN A $\boldsymbol{SU(2)_L}$ GAUGE THEORY
OF WEAK INTERACTIONS

AS A RESONANT EFFECT

OF 1-LOOP  TRANSITIONS BETWEEN QUASI-DEGENERATE FERMIONS}
\end{center}
}
\vskip .2cm
\centerline{ B.~Machet
     \footnote[1]{LPTHE tour 24-25, 5\raise 3pt \hbox{\tiny \`eme} \'etage,
          UPMC Univ Paris 06, BP 126, 4 place Jussieu,
          F-75252 Paris Cedex 05 (France),\\
         Unit\'e Mixte de Recherche UMR 7589 (CNRS / UPMC Univ Paris 06)}
    \footnote[2]{machet@lpthe.jussieu.fr}
     }
\vskip .7cm

{\bf Abstract:} We show that
1-loop transitions between two quasi-degenerate fermions can induce
a potentially large renormalization of their mixing angle, and a large
renormalized Cabibbo (or PMNS) angle when 
the second fermion pair in the same two generations is far
from degeneracy. At the resonance,  the ``Cabibbo angle'' gets maximal
and simply connected to masses without invoking any new physics beyond
the standard model. This solution appears as the only one ``perturbatively
stable'' (mixing angles are then renormalized
with respect to their classical values by small amounts).

\smallskip

{\bf PACS:} 12.15.Ff \quad 12.15.Lk \quad 14.60.Pq  \hfill

\section{Introduction}

The origin of large mixing angles observed in leptonic charged currents
is still unknown \cite{Smirnov}. A common idea is that it is linked to a
quasi-degeneracy of neutrinos, but this connection was never firmly
established. And it cannot be on simple grounds, since homographic
transformations on a (mass) matrix, while changing its eigenvalues,
 do not change its eigenvectors, hence mixing angles; accordingly,
infinitely different spectra can be associated to a given mixing angle.

We show below, in the case of binary coupled systems, that large mixing 
can be associated with quasi-degeneracy.
Indeed,  small (perturbative) changes of  parameters (for examples elements
of mass matrix) can then trigger large variations of  eigenstates. In the
case under scrutiny, 1-loop transitions between two  fermions
generate perturbative ${\cal O}(g^2)$ modifications of their kinetic terms.
A (slightly non-unitary) transformation,
which differs from a rotation only at ${\cal O}(g^2)$,
is needed to cast them back into their canonical form $\overline \Psi\,
\psl\, {\mathbb I}\, \Psi$ ($\mathbb I$ is the unit matrix).
When the two fermions are quasi-degenerate, the induced transformation
of their mass matrix is enough
to  trigger in turn large variations of its eigenvectors, such that its
re-diagonalization requires a rotation by a large angle. The latter
ultimately
becomes the renormalized Cabibbo angle that occurs in charged current.

In the following, we shall work with two generations of fermions,
and  take the example of two pairs of quarks $(d,s)$ and
$(u,c)$. This can be easily translated to the (more appropriate)
 lepton case,
when the two  pairs are instead, for example, $(\nu_e, \nu_\mu)$ and $(e^-, \mu^-)$.
Then, ``Cabibbo angle'' \cite{Cabibbo} translates into ``first PMNS angle''
$\theta_{12}$ \cite{PMNS}, ``quasi-degenerate
$(d,s)$ system'' into ``quasi-degenerate neutrino pair'', $(u,c)$ far from
degeneracy into $(electron, muon)$ far from degeneracy {\em etc}.
Also, the the sake of simplicity, we shall work in a pure $SU(2)_L$ theory
of weak interactions instead of the standard $SU(2)_L \times U(1)$
electroweak model.

\section{1-loop transitions between non-degenerate fermions.
Re-diagonalizing the quadratic Lagrangian}
\label{section:fermions}

\subsection{1-loop transitions}

Like in the Standard Model of electroweak interactions \cite{GSW},
 the diagonalization of the classical mass matrix by
a bi-unitary transformation leads to the classical mass eigenstates, for
example $s^0_m$ and $d^0_m$, with classical masses $m_s$ and $m_d$.
They are orthogonal  with respect to the classical Lagrangian
 (no transition between them occurs at the classical level).
However, at 1-loop, gauge interactions induce  diagonal and non-diagonal
transitions between them.  For example, Fig.~1 describes non-diagonal
$s^0_m \to d^0_m$ transitions
\footnote{Similar transitions occur between $c$ and $u$ quarks,
and between their leptonic equivalent.},
mediated by the $W^\pm$ gauge bosons. Diagonal transitions are mediated either
by $W_\mu^\pm$ or by  the $W_\mu^3$ gauge bosons.

\vbox{
\begin{center}
\includegraphics[height=4truecm,width=8truecm]{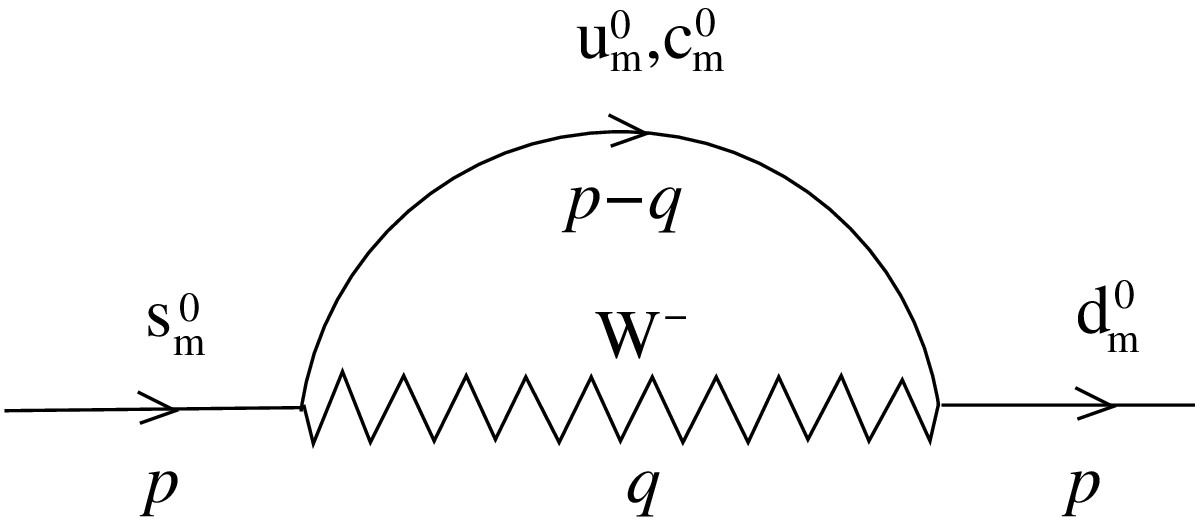}
\vskip 0pt
{\em Fig.~1: $s^0_m \to d^0_m$ transition at 1-loop}
\end{center}
}

We investigate in this work how the Cabibbo procedure
implements in the presence of these transitions \cite{shortDuMaVy}.
The one depicted in Fig.~1 contributes as a left-handed, kinetic-like, 
$p^2$-dependent interaction  of the type 
\begin{equation}
\sin\theta_c \cos\theta_c \big(h(p^2, m_u, m_W)-h(p^2, m_c,
m_W)\big) \, \bar d^0_m\, \psl (1-\gamma_5)\, s^0_m,
\label{eq:1loop}
\end{equation}
that we abbreviate, with transparent notations, into
\begin{equation}
s_c c_c (h_u-h_c) \, \bar d^0_m\; \psl\, (1-\gamma_5)\, s^0_m.
\label{eq:1loopbis}
\end{equation}
It depends on the classical Cabibbo angle
$\theta_c = \theta_d - \theta_u$.
The function $h$ is
dimensionless. It is simple matter to realize that  all
(diagonal and non-diagonal) 1-loop transitions  mediated  between $s$ and
$d$ mediated by $W^\pm$ gauge bosons  transform
their kinetic terms into
\begin{eqnarray}
&&\left(\begin{array}{cc} \bar d^0_m & \bar s^0_m\end{array}\right)
\left[ {\mathbb I}\; \psl +
\left(\begin{array}{cc}
c_c^2 h_u + s_c^2 h_c & s_c c_c(h_u -h_c) \cr
s_c c_c (h_u - h_c) & s_c^2 h_u + c_c^2 h_c
\end{array}\right)
\psl (1-\gamma_5)\right]
\left(\begin{array}{c} d^0_m \cr s^0_m
\end{array}\right)\cr
= && \left(\begin{array}{cc} \bar d^0_m & \bar s^0_m\end{array}\right)
\left[ {\mathbb I}\; \psl +
\left(\frac{h_u+h_c}{2} + (h_u-h_c)\; {\cal T}_x(2\theta_c)\right)
\psl (1-\gamma_5)\right]
\left(\begin{array}{c} d^0_m \cr s^0_m
\end{array}\right),
\label{eq:Wcont}
\end{eqnarray}
where we noted
\begin{equation}
{\cal T}_x(\theta) = \frac12\left(\begin{array}{rr}
\cos \theta & \sin\theta \cr \sin\theta & -\cos\theta \end{array}\right).
\label{eq:Tx}
\end{equation}
To the contributions (\ref{eq:Wcont}) we must add the diagonal
transitions mediated by the  $W^3_\mu$ gauge boson.
The kinetic terms for left-handed $d^0_m$ and $s^0_m$ quarks
become (omitting the fermionic fields)

\vbox{
\begin{eqnarray}
K_d &=& {\mathbb I} + H_d\; ;\cr
H_d &=& \frac{h_u+h_c}{2} + (h_u-h_c)\; {\cal T}_x(2\theta_c)
+ \left(\begin{array}{cc}h_d & \cr & h_s\end{array}\right),
\label{eq:Kd3}
\end{eqnarray}
}

where $h_d = h(p^2, m_d, m_W)$ and $h_s = h(p^2, m_s, m_W)$.
Likewise, in the $(u,c)$ sector, one has

\vbox{
\begin{eqnarray}
K_u &=& {\mathbb I} + H_u\; ;\cr
H_u &=& \frac{h_d+h_s}{2} + (h_d-h_s)\; {\cal T}_x(2\theta_c)
+ \left(\begin{array}{cc}h_u & \cr & h_c\end{array}\right).
\label{eq:Ku3}
\end{eqnarray}
}

We shall now diagonalize the quadratic part of the
 effective 1-loop Lagrangian, which
means putting the pure kinetic terms back to the unit matrix and,
at the same time, re-diagonalizing the mass matrix.
This is accordingly a two-steps procedure.

Note that the kinetic terms of right-handed fermions are not modified,
such that we shall only be concerned below  with the left-handed ones.

\subsection{First step: re-diagonalizing  kinetic terms 
back to the unit matrix}
\label{subsec:diakin}

The pure kinetic terms $K_d$  for $(d^0_m, s^0_m)$ written
in (\ref{eq:Kd3}) can be cast back to their canonical form by
a $p^2$-dependent non-unitary transformations  ${\cal V}_d$
according to
\begin{equation}
{\cal V}_{d}^\dagger\; K_d\; {\cal V}_{d} = {\mathbb I}.
\label{eq:kincon}
\end{equation}
The procedure to find ${\cal V}_d$ is the following.
Let $(1+t_+)$ and $(1+t_-)$, $t_+, t_- = {\cal O}(g^2)$,
 be the eigenvalues of the symmetric matrix
 $K_d$
\footnote{One has explicitly
\begin{equation}
t_\pm = \frac{h_u + h_c + h_d + h_s}{2}
\pm \frac12 \sqrt{\left(h_u - h_c\right)^2 + \left(h_d -
h_s\right)^2 
+ 2\;(h_u - h_c)\;(h_d - h_s)\cos 2\theta_c}.
\label{eq:t+-}
\end{equation}
.}.
It can be diagonalized by a rotation ${\cal
R}(\omega_d) \equiv\left(\begin{array}{rr} \cos\omega_d & \sin\omega_d \cr
-\sin\omega_d & \cos\omega_d
\end{array}\right)$ according to
\begin{equation}
{\cal R}(\omega_d)^\dagger\, K_d\, {\cal R}(\omega_d) = \left(\begin{array}{cc}
1+t_+ & \cr &  1+t_-\end{array}\right),
\label{eq:diag1}
\end{equation}
with
\begin{equation}
\tan 2\omega_d = \displaystyle\frac{-(h_u -h_c)\sin
2\theta_c}{(h_u-h_c)\cos 2\theta_c + h_d - h_s}
= \frac{-1}{1 + \frac{h_d - h_s}{h_u-h_c}}\; \tan 2\theta_c.
\label{eq:tanomega}
\end{equation}
(\ref{eq:tanomega}) defines $\omega_d$ in particular as a function of the
classical $\theta_c$: $\omega_d = \omega_d(\theta_c,\ldots)$.

The diagonal matrix obtained in (\ref{eq:diag1}) is not yet the unit matrix, but
one gets to it by a simple renormalization of the columns of ${\cal
R}(\omega_d)$ respectively by $\frac{1}{\sqrt{1+t_+}}$ and
$\frac{1}{\sqrt{1+t_-}}$. 
The looked for non-unitary matrix ${\cal V}_d$  writes finally
\begin{equation}
{\cal V}_d(p^2,\ldots) = \left(\begin{array}{rr}
\displaystyle\frac{c_{\omega_d}}{\sqrt{1+t_+}} &
\displaystyle\frac{s_{\omega_d}}{\sqrt{1+t_-}} \cr
-\displaystyle\frac{s_{\omega_d}}{\sqrt{1+t_+}} &
\displaystyle\frac{c_{\omega_d}}{\sqrt{1+t_-}}
\end{array}\right).
\label{eq:calVd}
\end{equation}
It differs from the rotation ${\cal R}(\omega_d)$ only at ${\cal O}(g^2)$
 and satisfies
\begin{equation}
{\cal V}_{d}\,{\cal V}_{d}^\dagger = 
\frac{1}{(1+t_+)(1+t_-)}\left( {\mathbb I} + \frac{t_++t_-}{2} -(t_+-t_-)
\;{\cal T}_x(-2\omega_d)\right), \quad
{\cal V}_d^\dagger\,{\cal V}_d = \left(\begin{array}{cc}
\displaystyle\frac{1}{1+t_+} & \cr & \displaystyle\frac{1}{1+t_-}
\end{array}\right).
\label{eq:VV}
\end{equation}
For  $|h_d -h_s| \ll |h_u -h_c|$, eq.~(\ref{eq:tanomega}) shows
that $\omega_d(\theta_c) \approx -\theta_c$. So,
when the pair $(d,s)$ is close to degeneracy and $(u,c)$ far from it
(see also footnote \ref{foot:maxi}),
${\cal V}_d$  becomes  close to a rotation ${\cal R}(-\theta_c)$.
This property plays, as shown in subsection \ref{subsec:rencab}, an important
role in the determination of the renormalized Cabibbo angle.

\subsection{Second step: re-diagonalization of the mass matrix}
\label{subsec:diamass}

By the flavor transformation ${\cal V}_d$ acting on left-handed
fermions in the bare mass basis, $M_d$ transforms into ${\cal V}_d^\dagger
M_d$, which needs to be re-diagonalized. To this purpose, a new bi-unitary
transformation is needed. The transformation acting on left-handed fermions
is the rotation ${\cal R}(\xi_d)$ that diagonalizes the real symmetric matrix
\footnote{From now onwards, to lighten the notations, we shall frequently omit the
dependence on $p^2$ and on the masses.}
\begin{eqnarray}
{\cal V}_d^\dagger\, M_d M_d^\dagger\, {\cal V}_d &=&
{\cal V}_{d}^\dagger
\left(\begin{array}{cc} m_d^2 & \cr & m_s^2\end{array}\right)
{\cal V}_{d}\cr
&=& \left(\begin{array}{cc}
\displaystyle\frac{m_d^2\, c_{\omega_d}^2 + m_s^2 s_{\omega_d}^2}{1+t_+} &
-\displaystyle\frac{s_{\omega_d} c_{\omega_d}(m_s^2-m_d^2)}{\sqrt{(1+t_+)(1+t_-)}} \cr
-\displaystyle\frac{s_{\omega_d} c_{\omega_d}(m_s^2-m_d^2)}{\sqrt{(1+t_+)(1+t_-)}} &
\displaystyle\frac{m_d^2 s_{\omega_d}^2 + m_s^2 c_{\omega_d}^2}{1+t_-}
\end{array}\right),
\label{eq:calM}
\end{eqnarray}
according to
\footnote{The re-diagonalization of kinetic terms indirectly contributes to
a renormalization of the masses: $m_d \to \mu_d, m_s \to \mu_s$.}
\begin{equation}
{\cal R}(\xi_d)^\dagger\,\left(
{\cal V}_d^\dagger\, M_d M_d^\dagger\, {\cal V}_d\right)
 \, {\cal R}(\xi_d) =
\left(\begin{array}{cc} \mu_d^2 & \cr & \mu_s^2 \end{array}\right).
\label{eq:diagM}
 \end{equation}
It satisfies 
\begin{equation}
\tan 2\xi_d = \displaystyle\frac
{-(m_d^2-m_s^2)\sqrt{(1+t_+)(1+t_-)}\sin 2\omega_d}
{(m_d^2-m_s^2)\left(1+\displaystyle\frac{t_+ +t_-}{2} \right)\cos 2\omega_d
- (m_d^2+m_s^2)\displaystyle\frac{t_+-t_-}{2}}.
\label{eq:xid}
\end{equation}
(\ref{eq:xid}) defines in particular $\xi_d$ as a function of $\omega_d$,
and thus as a function of the classical $\theta_c$: $\xi_d =
\xi_d(\theta_c,\ldots)$.
It also defines 1-loop mass eigenstates $d_{mL}(p^2)$ and $s_{mL}(p^2)$.
Since it is in particular unitary, it preserves the canonical form of the
kinetic terms that had been recovered in the first step of the procedure.
By construction, at any given $p^2$, there is no 1-loop transition between
$d_{mL}(p^2)$ and $s_{mL}(p^2)$.

The main property of (\ref{eq:xid}) is the presence of a pole. 
It occurs for
\begin{equation}
2\;\displaystyle\frac{m_d^2-m_s^2}{m_d^2+m_s^2}\, \cos 2\omega_d(\theta_c)
\approx t_+ - t_-
\stackrel{(\ref{eq:t+-})}{=}   \sqrt{\left(h_u - h_c\right)^2 + \left(h_d -
h_s\right)^2 
+ 2\;(h_u - h_c)(h_d - h_s)\,\cos 2\theta_c},
\label{eq:pole}
\end{equation}
which is, through (\ref{eq:tanomega}), a  relation between
$\theta_c$, $m_d,m_s,m_u,m_c$, $m_W$ (and $p^2$).

We shall see in subsection \ref{subsec:rencab} that, for quasi-degenerate
$(d,s)$ and largely split $(u,c)$, $\xi_d(\theta_c)$ ultimately becomes the
renormalized Cabibbo angle, which is accordingly implicitly
expressed by (\ref{eq:xid}) as
a function of the masses of fermions and gauge fields, and of $p^2$.

\section{Individual mixing matrix and  renormalized mixing angle}

\subsection{1-loop and classical mass eigenstates are non-unitarily
related} \label{subsec:non-unit}

The left-handed
\footnote{The subscript ${_L}$ refers to left-handed fermions.} 
1-loop mass eigenstates are related to the bare ones by
\begin{equation}
\left(\begin{array}{c} d^0_{mL} \cr s^0_{mL}\end{array}\right)
= {\cal V}_d\, {\cal R}(\xi_d) \left(\begin{array}{c}
d_{mL} \cr s_{mL} \end{array}\right).
\label{eq:trans}
\end{equation}
They are thus deduced from the latter by the product of a $p^2$-dependent
non-unitary transformation
${\cal V}_{d}$ and a $p^2$-dependent unitary one ${\cal R}(\xi_d)$.
The two basis are accordingly non-unitarily related \cite{mixing}.
In particular, on mass-shell (respectively at $p^2=m_d^2$ and $p^2 = m_s^2$),
one has for the physical mass eigenstates
\begin{eqnarray}
d_{mL}^{phys} \equiv d_{mL}(p^2 = m_d^2) &=& [{\cal V}_d\,{\cal R}(\xi_d)]_{11}(p^2=m_d^2)\; d^0_{mL} + [{\cal
V}_d\,{\cal
R}(\xi_d)]_{12}(p^2=m_d^2)\; s^0_{mL}, \cr
s_{mL}^{phys} \equiv s_{mL}(p^2 = m_s^2) &=& [{\cal V}_d\,{\cal R}(\xi_d)]_{21}(p^2=m_s^2)\; d^0_{mL} + [{\cal
V}_d\,{\cal R}(\xi_d)]_{22}(p^2=m_s^2)\; s^0_{mL}.
\label{eq:nonor}
\end{eqnarray}
Since bare mass states are unitarily related to bare flavor states, 
the physical mass eigenstates  are also non-unitarily
related to bare flavor states.

\subsection{Individual mixing matrix and renormalized mixing angle}
\label{subsec:indiv}

Classical flavor eigenstates and 1-loop mass eigenstates are related to
each other according to
\begin{eqnarray}
\left(\begin{array}{c} d^0_{fL} \cr s^0_{fL}\end{array}\right)
= {\cal C}_{d0} \left(\begin{array}{c}
d^0_{mL} \cr s^0_{mL} \end{array}\right)
\stackrel{(\ref{eq:trans})}{=}
 {\cal C}_{d0}\,{\cal V}_d\, {\cal R}(\xi_d) \left(\begin{array}{c}
d_{mL} \cr s_{mL} \end{array}\right),
\end{eqnarray}
where ${\cal C}_{d0} \equiv {\cal R}(\theta_d)$
 is the classical mixing matrix in the $(d,s)$ sector.
The individual mixing  matrix at 1-loop is thus given by
\begin{equation}
{\cal C}_{d} = {\cal C}_{d0}\,{\cal V}_{d}\,{\cal R}(\xi_d)
= {\cal R}(\theta_d)\, {\cal V}_d\, {\cal R}(\xi_d).
\label{eq:Cd}
\end{equation}

Since ${\cal V}_d \approx {\cal R}(\omega_d) + {\cal O}(g^2)$ (see
(\ref{eq:calVd})), one has
\begin{equation}
{\cal C}_d
\approx {\cal R}(\theta_d +\omega_d +\xi_d).
\label{eq:indiv}
\end{equation}
The quantity $\omega_d + \xi_d$ is seen on (\ref{eq:indiv}) to renormalize
the classical mixing angle $\theta_d$.
From (\ref{eq:xid}), one deduces that it satisfies the general relation
\begin{equation}
\tan 2(\omega_d +\xi_d) \approx\frac
{-\tan 2\omega_d\left[\frac{t_+-t_-}{2}\;\frac{m_d^2+m_s^2}{m_d^2-m_s^2}
\;\frac{1}{\cos2\omega_d}\right]}
{1 + \tan^2 2\omega_d
-\left[\frac{t_+-t_-}{2}\;\frac{m_d^2+m_s^2}{m_d^2-m_s^2}\;\frac{1}{\cos2\omega_d}\right]}.
\label{eq:xiomega}
\end{equation}
Let us suppose now that $d$ and $s$ are quasi-degenerate and that
 $u$ and $c$ are, at the opposite far from degeneracy.  Then
(see subsection \ref{subsec:diakin}), $\omega_d(\theta_c)
\approx -\theta_c$, and (\ref{eq:indiv}) becomes
 ${\cal C}_d \approx {\cal R}(\theta_d -\theta_c +\xi_d(\theta_c))
 = {\cal R}(\theta_u +\xi_d(\theta_c))$.
Furthermore, at the pole (\ref{eq:pole}) of (\ref{eq:xid}),
{\em i.e.} when $\xi_d(\theta_c)$ becomes maximal $\xi_d = \pm\pi/4$,
  it is easy to show that  $\omega_d + \xi_d$ gets  small
\footnote{When the pair $(d,s)$ is quasi-degenerate and $(u,c)$
far from degeneracy,
 $\omega_d \stackrel{(\ref{eq:tanomega})}{\approx} -\theta_c$
such that (\ref{eq:pole}) is approximately a second
degree equation in $\cos 2\theta_c$. Furthermore, one has, then (in the
unitary gauge),
$|h_d -h_s| \stackrel{m_d^2,m_s^2,p^2 \ll m_W^2}{\approx}
 \frac{g^2}{8\pi^2}\frac{m_s^2 - m_d^2}{m_W^2} \ll
|h_u -h_c| \stackrel{m_u^2,m_c^2,p^2 \ll m_W^2} {\approx}
\frac{g^2}{8\pi^2}\frac{m_c^2 - m_u^2}{m_W^2}$,
which, added to $|h_d -h_s| \ll
\frac{m_s^2-m_d^2}{m_s^2+m_d^2}$, enables to write the
approximate solution of (\ref{eq:pole}) as
\begin{equation}
\cos 2\theta_c \approx \frac12 (h_u -
h_c)\;\frac{m_d^2+m_s^2}{m_d^2-m_s^2}
\approx\frac{g^2}{16\pi^2}\;\frac{m_c^2-m_u^2}{m_W^2}
\;\frac{m_s^2+m_d^2}{m_s^2-m_d^2}.
\label{eq:pole2}
\end{equation}
Since the r.h.s of (\ref{eq:pole2}) is $\ll 1$, it corresponds to
 a classical $\theta_c$ itself close to maximal. Then, so does
$\omega_d(\theta_c)$.

At the pole (\ref{eq:pole}),
$\left[\frac{t_+-t_-}{2}\;\frac{m_d^2+m_s^2}{m_d^2-m_s^2}\;\frac{1}{\cos
2\omega_d}\right]=1$ and the relation (\ref{eq:xiomega}) becomes
$\tan 2(\omega_d+\xi_d) = -1/\tan 2\omega_d$,  which
vanishes when $\omega_d$ becomes maximal. Then,
 $\omega_d + \xi_d \to 0$, {\em q.e.d.}
\label{foot:maxi}}
; $\theta_d$ is then
renormalized only by a small amount
\footnote{Ones finds numerically from (\ref{eq:xiomega}) and
(\ref{eq:tanomega}) that $(\omega_d + \xi_d)(\theta_c)$ only vanishes at
the pole (\ref{eq:pole}), {\em i.e.} when $\theta_c \approx -\omega_d$
is close to maximal.  \label{foot:omegaxi}}.

\section{The renormalized Cabibbo  angle}
\label{section:rencab}

\subsection{The effective gauge-invariant and hermitian
Lagrangian at 1-loop}
\label{subsec:Leff}

After 1-loop radiative corrections to $s^0_{mL} \leftrightarrow d^0_{mL}$
and $c^0_{mL} \leftrightarrow u^0_{mL}$ have been accounted for,
the kinetic terms for the first two  generations of left-handed
fermions, once cast into their standard form
$\overline\Psi\, \overleftrightarrow D \Psi \equiv
\frac{1}{2} \big(\overline\Psi D \Psi - (\overline{D \Psi})\Psi
\big)$, write, in the bare mass basis
\begin{equation}
\hskip -1cm{\cal L} \in
\left(\begin{array}{cccc} \bar u^0_{mL} & \bar c^0_{mL} & \bar
d^0_{mL}  &\bar s^0_{mL}\end{array}\right)
\left( A\, \psl -\frac{ig}{2} (A\, \vec T + \vec T A). \vec W_\mu)\, \gamma^u + \ldots \right)
\left(\begin{array}{c} u^0_{mL} \cr c^0_{mL} \cr d^0_{mL} \cr s^0_{mL}
\end{array}\right),
\label{eq:Lcharged}
\end{equation}
with
\begin{equation}
A =
 \left(\begin{array}{ccc}
K_u & \vline &   \cr \hline  & \vline & K_d \end{array}\right)
={\mathbb I} + \left(\begin{array}{ccc}
H_u & \vline &   \cr \hline  & \vline & H_d \end{array}\right).
\label{eq:kcterm}
\end{equation}
$SU(2)_L$ gauge invariance, by requesting the replacement of the partial
derivative by the covariant one, is at the origin of the gauge couplings
that appear in (\ref{eq:Lcharged}).
${\cal L}$ is hermitian and involves the (Cabibbo rotated) $SU(2)_L$
generators $\vec T$
\begin{equation}
T^3 = \frac12
\left(\begin{array}{ccc}
1   & \vline &   \cr
\hline
& \vline & -1    \end{array}\right),
T^+ = \left(\begin{array}{ccc}
 &   \vline &  {\cal C}_{0} \cr
\hline
& \vline & \end{array}\right),
T^- = \left(\begin{array}{ccc}
 &   \vline   &  \cr
\hline
{\cal C}^\dagger_{0} & \vline &  \end{array}\right),
\label{eq:TTT}
\end{equation}
where ${\cal C}_0$ is the classical Cabibbo matrix
\begin{equation}
{\cal C}_0 = {\cal R}(\theta_c)
= \left(\begin{array}{rr}
\cos\theta_c & \sin\theta_c \cr -\sin\theta_c & \cos\theta_c
\end{array}\right)
 = {\cal C}_{u0}^\dagger\, {\cal C}_{d0} =
{\cal R}(\theta_u)^\dagger\, {\cal R}(\theta_d).
\label{eq:bareCab}
\end{equation}

\subsection{The renormalized Cabibbo angle}
\label{subsec:rencab}

From (\ref{eq:Lcharged}), one deduces that, in the bare mass basis, the
renormalized Cabibbo matrix is, at ${\cal O}(g^2)$

\begin{equation}
{\cal C}^{bm}(p^2,\ldots) =
\frac12\big[
 \underbrace{({\mathbb I} + H_u)}_{K_u}\, {\cal C}_{0}
+{\cal C}_{0}\,\underbrace{({\mathbb I} + H_d)}_{K_d}\big]
\label{eq:Cab2}
\end{equation}
which, in particular,  is not unitary
.
Using (\ref{eq:trans}), it becomes in the basis of 1-loop mass eigenstates
\begin{eqnarray}
 {\mathfrak C}(p^2,\ldots) = [{\cal V}_u\, {\cal R}(\xi_u)]^\dagger\, {\cal
C}^{bm}(p^2,\ldots)\,
[{\cal V}_d\, {\cal R}(\xi_d)],
\label{eq:cabrenor}
\end{eqnarray}
Since $H_{u}$ and $H_d$ in (\ref{eq:Cab2}) are ${\cal O}(g^2)$, the
terms proportional to them in (\ref{eq:cabrenor}) can be calculated with
the expressions of ${\cal R}(\xi_d)$ and ${\cal V}_d$ at ${\cal O}(g^0)$,
that is, for $t_+ = 0 =t_-$; one can accordingly take in there
${\cal R}(\xi_d) \stackrel{(\ref{eq:xid})}{\to} {\cal R}(-\omega_d)$
and ${\cal V}_d \stackrel{(\ref{eq:calVd})}{\to} {\cal R}(\omega_d)$, such
that ${\cal V}_d {\cal R}(\xi_d) \to {\mathbb I}$. The
same approximation can be done in the $(u,c)$ sector.
 The resulting expression for $\mathfrak C$ is 
\begin{eqnarray}
 {\mathfrak C}(p^2,\ldots) &\stackrel{{\cal O}(g^2)}{\approx}&
{\cal R}(\xi_u)^\dagger\, {\cal V}_u^\dagger\;\,
{\cal C}_0\; {\cal V}_d\, {\cal R}(\xi_d)
+ \frac12 \big( H_u\, {\cal C}_0 + {\cal C}_0\, H_d \big)\cr
&=& {\cal C}_u^\dagger\, {\cal C}_d + {\cal O}(g^2),
\label{eq:cabrenor2}
\end{eqnarray}
in the second line of which we have used (\ref{eq:bareCab}),
 (\ref{eq:Cd}) and its equivalent for ${\cal C}_u$.

Let us now get an approximate expression for ${\cal C}_u^\dagger\,
 {\cal C}_d$ when $d$ and $s$ are close to degeneracy,
while $u$ and $c$ are far from it. Then (see subsection
\ref{subsec:diakin}),
$\omega_d(\theta_c) \approx -\theta_c$ such that, by (\ref{eq:calVd}), one has
 ${\cal V}_d \approx {\cal R}(-\theta_c)$, which cancels the
${\cal C}_0 \equiv {\cal R}(\theta_c)$ in (\ref{eq:cabrenor2}).
Likewise, from the equivalent
$\tan 2\omega_u = \frac{-(h_d -h_s)\sin
2\theta_c}{(h_d-h_s)\cos 2\theta_c + h_u - h_c}$
 of (\ref{eq:tanomega}) in the $(u,c)$
sector, we deduce that, since $|h_u - h_c| \gg |h_d-h_s|$,
 $\omega_u \to 0$ such that, from the equivalent of
(\ref{eq:calVd}),  ${\cal V}_u \approx {\mathbb I}$.
Also, since $\tan 2\xi_u$ is proportional to $\sin 2\omega_u$ in the equivalent
of (\ref{eq:xid}), $\xi_u$ becomes  small, such that ${\cal
R}(\xi_u) \to {\mathbb I}$, too
\footnote{${\cal V}_u{\cal R}(\xi_u) \approx 1$,
such that the renormalization of the mixing angle of the largely
split pair is small.\label{foot:thetau}}.
Finally, (\ref{eq:cabrenor2}) becomes
\begin{equation}
{\mathfrak C}(p^2,\ldots) \approx {\cal R}(\xi_d(\theta_c)) + {\cal O}(g^2)\ \text{when}\ 
(d,s) \approx\ \text{degenerate and}\ (u,c)\ \text{far from degeneracy}.
\end{equation}
This is our main result: the renormalized value of
the Cabibbo angle finally becomes $\xi_d(\theta_c)$ as given by
(\ref{eq:xid}); it can become large and eventually maximal
at the resonance (\ref{eq:pole}).
If so, since $\theta_c$ is then close to maximal, too
(see footnote \ref{foot:maxi}), the Cabibbo angle gets renormalized
by a small amount (like $\theta_d$ (see subsection \ref{subsec:indiv})
 and $\theta_u$ (see footnote \ref{foot:thetau})).

\section{Summary and prospects}

We have shown that, in a $SU(2)_L$ gauge model  of weak interactions,
1-loop transitions between two fermions can strongly modify their mass
eigenstates and generate a large mixing angle when:\newline
* this pair  is close to degeneracy;\newline
* the other pair in the same two generations is, at the opposite,
 far from degeneracy.\newline
While the classical mixing angle $\theta_u$ of the largely split
pair undergoes a small renormalization, 
the one $\theta_d$ of the quasi-degenerate pair gets renormalized by
$\xi_d(\theta_c) -\theta_c$, which play the following roles:
the rotation ${\cal R}(\theta_c)$  casts 
the kinetic terms of the quasi-degenerate pair back to the unit matrix
and ${\cal R}(\xi_d(\theta_c))$ puts its mass matrix back  to diagonal.
The Cabibbo angle gets accordingly  renormalized
from $\theta_c \equiv\theta_d - \theta_u$ to, approximately,
$(\theta_d + \xi_d(\theta_c) -\theta_c) -\theta_u$,
that is, up to corrections ${\cal O}(g^2)$, $\xi_d(\theta_c)$ itself.
In the vicinity of the pole of $\tan 2\xi_d$, both $\theta_c$ and $\xi_d$
become close to maximal. 1-loop renormalizations
of $\theta_d$,  $\theta_c$ and  $\theta_u$ are then small.
A maximal value for the Cabibbo angle appears in these conditions
as the only perturbatively stable solution (see footnote
\ref{foot:omegaxi}).

This  non-trivial effect of 1-loop radiative corrections
could explain the large mixing angles
observed in charged leptonic currents if the classical PMNS
angles are close to fulfilling the leptonic equivalent of
conditions (\ref{eq:pole}) and (\ref{eq:pole2}).
To our knowledge, it is the first time that such relations
connecting masses and angles
could be established on simple perturbative grounds without invoking
physics beyond the standard model.

A more quantitative analysis is currently under investigation.

\vskip .6cm 
\begin{em}
\underline {Acknowledgments}: It is a pleasure to thank M.I. Vysotsky for
comments and advice.
\end{em}
\vfill
\begin{em}

\end{em}

\end{document}